\documentclass[graybox]{svmult}

\usepackage{mathptmx}       
\usepackage{helvet}         
\usepackage{courier}        
\usepackage{type1cm}        
\usepackage{makeidx}         
\usepackage{graphicx}        
\usepackage{multicol}        
\usepackage[bottom]{footmisc}

\usepackage{amsmath}

\makeindex             

\newcommand{\beq}{\begin{equation}}
\newcommand{\eeq}{\end{equation}}
\newcommand{\bea}{\begin{eqnarray}}
\newcommand{\eea}{\end{eqnarray}}

\newcommand{\Mbh}{M_{\rm BH}}
\newcommand{\Mpl}{M_{\rm Pl}}

\begin{document}

\title*{Self-Completeness in Alternative Theories of Gravity}
\author{Maximiliano Isi, Jonas Mureika and Piero Nicolini}
\institute{Maximiliano Isi \at Loyola Marymount University, Los Angeles, CA 90045-2659,
 \email{misi@lion.lmu.edu}
\and Jonas Mureika\at Loyola Marymount University, Los Angeles, CA 90045-2659,
 \email{jmureika@lmu.edu}
\and Piero Nicolini \at Frankfurt Institute for Advanced Studies, Ruth-Moufang-Strasse 1 60438, Frankfurt am Main, and Institut f\"{u}r Theoretische Physik, J. W. Goethe-Universit\"{a}t, Max-von-Laue-Strasse 1, 60438 Frankfurt am Main, Germany,
 \email{nicolini@fias.uni-frankfurt.de}}
%
%
\maketitle

\abstract*{t has recently been shown via an equivalence of gravitational radius and Compton wavelength in four dimensions that the trans-Planckian regime of gravity may by semi-classical, and that this point is defined by a minimum horizon radius commensurate with the Planck mass. We extend the formalism to modified theories of gravity to the formalisms of Randall-Sundrum and the generelized uncertainty principle.}

\abstract{It has recently been shown via an equivalence of gravitational radius and Compton wavelength in four dimensions that the trans-Planckian regime of gravity may by semi-classical, and that this point is defined by a minimum horizon radius commensurate with the Planck mass. We extend the formalism to modified theories of gravity to the formalisms of Randall-Sundrum and the generelized uncertainty principle.}

\section{Introduction}
From the perspective of quantization, gravity is problematic since, among other reasons, graviton path integrals in $(3+1)$-D are readily divergent. As part of the effort to solve this, it has been shown \cite{dvali1,dvali2,dvali3,euro1,eurosmail} that gravity may be considered ``self-complete,'' in that there exists a minimum horizon scale hiding the singularity.  Specifically, this distance is defined by the confluence of the classical Schwarzschild radius and the Compton wavelength,
\beq
r_H = \lambda_C ~~~\Longrightarrow~~~\frac{2G\Mbh}{c^2} = \frac{h}{ c \Mbh}~.
\eeq
Letting $\Mpl=\sqrt{\hbar c/G}$, this gives a minimum mass:
\beq
\Mbh \geq \sqrt{\frac{h c}{2G}} =\sqrt{\pi}\Mpl~,
\label{eq:selfcomplete}
\eeq
below which the effective length scale increases as $\Mbh^{-1}$.  That is, Planck-mass holes are the smallest resolvable black objects.

\section{Randall-Sundrum}
The Randall-Sundrum model posits our Universe is a $n$-dimensional brane in a bulk with an infinite extra dimension at a distance $\ell$, the AdS curvature radius \cite{rm} .  Einstein's equations in the bulk are
\beq
\tilde{G}_{AB} = \tilde{\kappa}^2 \left[ -\tilde{\Lambda} \tilde{g}_{AB} + \delta(\chi) \left(-\lambda g_{AB} + T_{AB}\right) \right]~,
\eeq
where the coupling $\tilde{\kappa} = 8\pi/ \tilde{\Mpl}^{3}$ is a function of the reduced $(n+1)$-dimensional Planck mass $\tilde{\Mpl}$.  The hierarchy problem is thus resolved by assuming that originates on the extra brane, causing our effective gravitational constant to be $G_4 = G_5 / \ell$, where $G_5$ is the ``true'' coupling strength \cite{reissner_metric}. 

\begin{figure} [t]
\sidecaption[t]
  \includegraphics[scale=.8]{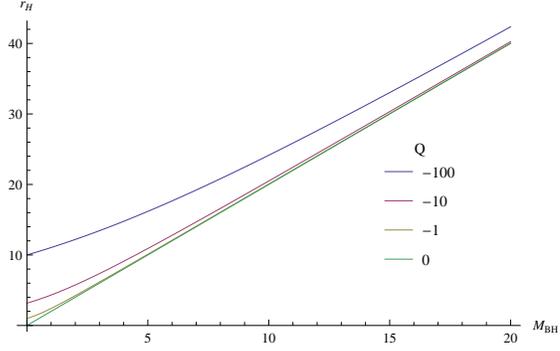}
  \caption{Black hole horizon radius as a function of mass in the Randall-Sundrum model (Planck units). The plot displays (\ref{eq:rRS}) for different values (negative) of the tidal charge Q. Note that the mass is in units of $\Mpl$.}
  \label{fig:rRS}
\end{figure}

In the case of an electrically neutral black hole, the induced Einstein equations on the brane yield a Reissner-Nordstr\"om-like solution of the form \cite{reissner_metric}\footnote{Note that \cite{reissner_metric} uses $\beta$ for the tidal charge and Q for the electric charge, which we take to be null.}
\beq \label{eq:metricRS}
ds_4^2 = -\left(1-\frac{2 G_4 m}{c^2 r} + \frac{Q}{r^2}\right) c^2 dt^2 + \frac{dr^2}{1-\frac{2 G_4 m}{c^2 r} + \frac{Q}{r^2}} + r^2 d\Omega^2~,
\eeq
with $d\Omega^2=d\theta^2+sin^2\theta d\phi^2$ and the term Q is the \emph{tidal charge}, resulting from leakage into the bulk. Knowing that $G_4=\hbar c /\Mpl^2$, we can write (\ref{eq:metricRS}) in terms of the Planck mass 
to obtain the (outer) black hole horizon:
\beq \label{eq:rRS}
r_H = \frac{\hbar}{c}\frac{\Mbh}{\Mpl^2}\left(1+\sqrt{1-\frac{c^2}{\hbar^2}\frac{\Mpl^4}{\Mbh^2}Q}\right)~.
\eeq
For the external horizon to be greater than the Schwarzschild radius, we require a negative $Q$. Otherwise, both radii would be smaller than the regular Schwarzschild horizon and we would get the usual self-completeness condition from (\ref{eq:selfcomplete}). Furthermore, $Q<0$ is arguably a more ``physical" choice \cite{dadhich}. Regardless of $Q$, $r_H(\Mbh)$ becomes linear for large enough mass (Fig. \ref{fig:rRS}).

Equating the Compton wavelength to the black hole horizon radius, we obtain an expression for the minimum black hole mass as a function of $Q$:
\beq \label{eq:MbhRS}
\Mbh \geq \frac{\pi \Mpl}{\sqrt{\pi-c^2\Mpl^2 Q/4\hbar^2}}
\eeq

\begin{figure} [t]
\sidecaption[t]
  \includegraphics[scale=.57]{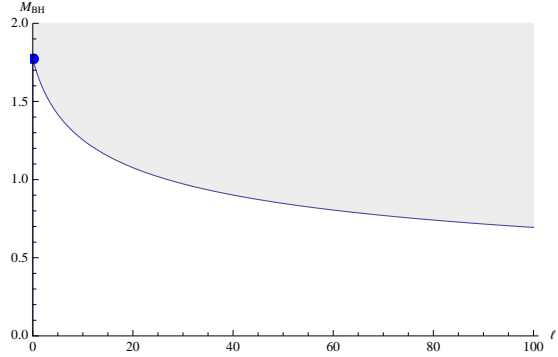}
  \caption{Minimum black hole mass (\ref{eq:scRS}) in Planck-mass units as a function of brane separation (solid line). The shaded region indicates the allowed values of the mass. As the correction is removed (viz. $\ell\rightarrow0^+$), the mimimum mass is again $\sqrt{\pi}\Mpl$ (indicated by a dot), agreeing with (\ref{eq:selfcomplete}).}
  \label{fig:RS_mL}
\end{figure}

However, $Q$ is not really an independent variable. For small length scales, compared to the AdS radius ($r\ll\ell$), the tidal charge becomes a linear function of the brane separation distance:
\beq Q \underset{r \ll \ell}{\approx} -\frac{M_{BH}}{\Mpl^{2}} \frac{\hbar}{c}\ell~.\eeq
Consequently, the minimum mass is also a function of $\ell$. In fact, after some basic algebra, we find
\beq \label{eq:scRS}
\Mbh \geq A \left(B + B^{-1} -1\right)~,
\eeq
\beq
A\equiv\frac{4\pi}{3}\frac{\hbar}{c\ell}~~,~~B\equiv\left[\frac{3\pi}{2}\left(\frac{\Mpl}{A}\right)^2 -1 +\frac{\Mpl}{A}\sqrt{\left(\frac{3\pi}{2}\frac{\Mpl}{A}\right)^2-3\pi}\right]^{1/3}~.
\eeq
The meaning of eq. (\ref{eq:scRS}) can be illuminated by means of a simple expansion:
\beq
M_{\rm min} = \sqrt{\pi} ~\Mpl - \frac{c \ell}{8 \hbar} ~\Mpl^2 + \frac{5}{128}\left(\frac{c \ell}{\hbar}\right)^2\pi^{-1/2}~ \Mpl^3 + {\cal O} (\Mpl^4)~.
\eeq
Again, we recover (\ref{eq:selfcomplete}) for vanishing $\ell$, as expected (Fig. \ref{fig:RS_mL}). On the other hand, note that  $M_{\rm min}\rightarrow 0$ as $\ell\rightarrow\infty$. Furthermore, because (\ref{eq:scRS}) is continous for all positive values of $\ell$ (which we require in order to have $Q<0$), Randall-Sundrum gravity can always be considered self-complete.

\section{Generalized Uncertainty Principle}

If additional momentum dependent terms exist in the usual commutation relation, this will result in a modified uncertainty relation of the form $\Delta x \Delta p \geq \frac{\hbar}{2} \left( 1 + \beta(\Delta p)^2  \right)$.  Such modification is known by the name of \emph{generalized uncertainty principle} (GUP). In turn, eq. this introduces a non-zero commutator between the coordinate operators:
\beq
[{\bf x}_i,{\bf x}_j] = 2i \hbar \beta ({\bf p}_i{\bf x}_j-{\bf p}_j{\bf x}_i)~.
\eeq
Because the commutator does not vanish unless $\beta=0$, the GUP introduces a non-zero minimal uncertainty in position, which translates into the existence of a minimal length. Furthermore, this results in a a momentum integration measure
\beq
\int{\frac{d^np}{1+\beta {\bf p}^2}| p \rangle \langle p |}=1~,
\eeq
which presents a UV cutoff of $\sqrt{\beta}$ \cite{kempf}. This has important consequences for black hole evaporation and results in remnant formation.

\begin{figure} [b]
\sidecaption[b]
  \includegraphics[scale=.8]{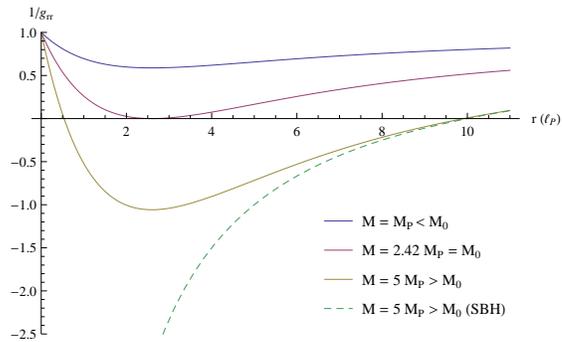}
  \caption{Metric coefficient for GUP-inspired metric (\ref{eq:GUP_metric}). Notice naked singularity, extremal and regular black hole cases. The Schwarzschild (SBH) case for $M=5 \Mpl$ is shown for comparison.}
  \label{fig:GUP_lapse}
\end{figure}
The GUP replaces the Dirac delta in the description of point particles of regular quantum mechanics with a wider Gaussian distribution, $e^{-|\vec{x}|\sqrt{\beta}}$. As shown in \cite{isi}, we can reproduce these non-local effects by means of the GUP-inspired metric
\beq \label{eq:GUP_metric}
ds^2=-\left(1-2\frac{GM}{c^2 r}\gamma(2;r/\sqrt{\beta})\right)dt^2-\left(1-2\frac{GM}{c^2 r}\gamma(2;r/\sqrt{\beta})\right)^{-1}dr^2+r^2d\Omega^2
\eeq
where $\gamma(s;x)=\int^x_0 t^{s-1} e^{-t}dt$ is the lower incomplete gamma function. The metric coefficient $1/g_{rr}$ is shown in Fig. \ref{fig:GUP_lapse}. Note that the extremal case happens at $\Mbh\approx 1.66 \sqrt{\beta}/G$ and $r_H\approx 1.73 \sqrt{\beta}$.

\begin{figure} [t]
\sidecaption[t]
  \includegraphics[scale=.75]{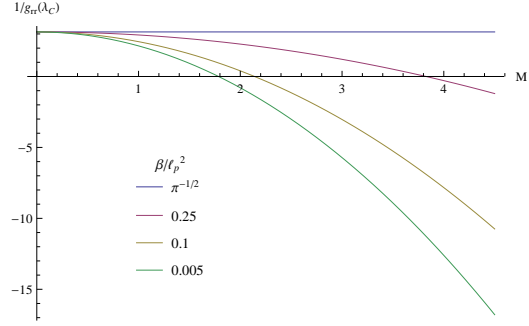}
  \caption{GUP auxiliary function (\ref{eq:GUP_aux}) for different values of minimum area $\beta$, shown Planck units. The roots indicate the minimum black hole mass for the particular value of $\beta_0$, where $\beta_0=\beta/\ell_P^2$. Note that there are no roots for $\beta_0\geq1/\pi$. This indicates that for large enough minimum areas, GUP stops being self-complete.}
  \label{fig:GUP_aux}
\end{figure}

It is not possible to find an explicit expression for the horizon radius corresponding to (\ref{eq:GUP_metric}. However, we can naively attempt to study the self-completeness of this metric by numerically solving $1/g_{rr}=0$ under the constraint $r=\lambda_{\rm C}(M)$, viz.
\beq \label{eq:GUP_aux}
1-2\frac{GM}{c^2 \lambda_{\rm C}}\gamma(2;\lambda_{\rm C}/\sqrt{\beta})=0
\eeq
Rather than taking the usual expression for $\lambda_{\rm C}$, we follow \cite{carr, adler_AJP} by  correcting the Compton wavelength to account for GUP effects: 
\beq \label{eq:GUP_lambda}
\lambda_{GUP}=\frac{\hbar}{M c}(1+ \beta M^2)~.
\eeq
Note that this step is not strictly required (see \cite{isi} for a more rigorous approach). The RHS of eq. (\ref{eq:GUP_gammaR}) is plotted in Fig.  \ref{fig:GUP_aux} The roots of this function can be interpreted as the values of $\Mbh$ at which the horizon radius coincides with the modified Compton wavelength for a given $\beta$, i.e. a minimum black hole mass.

\begin{figure}[t]
\sidecaption[t]
  \includegraphics[scale=.56]{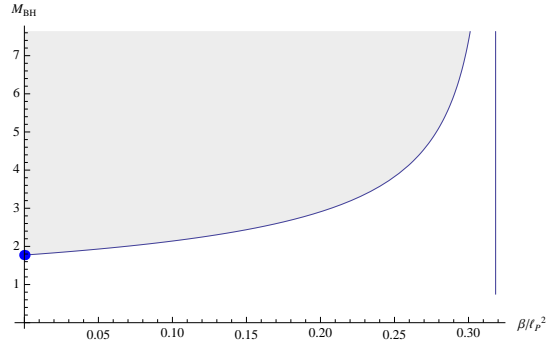}
  \caption{Black hole mass in Planck units for varying $\beta$. The allowed values correspond to the shaded region. The minimum-mass curve (solid), which was obtained numerically, presents an asymptote at $\beta=\ell_P^2/\pi\approx0.318$. For $\beta=0$, we recover the GR constraint (dot).}
  \label{fig:GUP_mb}
\end{figure}

The allowed black hole masses are shown in Fig.\ref{fig:GUP_mb}. The GUP corrections can be undone by letting $\beta \rightarrow 0$, thus recovering eq. (\ref{eq:selfcomplete}). Furthermore, we find that (\ref{eq:GUP_aux}) has no positive roots for $\beta/\ell_{\rm P}^2\geq1/\pi\approx0.318$, where $\ell_P=\sqrt{\hbar G / c^3}$ is the Planck length. Consequently, GUP is not self-complete for $\beta\geq \ell_P^2/\pi$. This can be turn into an upper bound on the minimum area: $\beta < \ell_P^2/\pi$.
Note that this is a constraint several orders of magnitude stronger than those found in \cite{Das} of $\beta_0<10^{21}$ and the corresponding energies are too high to be tested with current experiments.

\section{Conclusions}
We have explored the self-completeness of gravity under two different and independent frameworks: Randall-Sundrum and GUP. In the case of Randall-Sundrum, we have shown that gravity should be self-complete regardless of the AdS curvature radius and found an closed-form solution for the minimum mass, eq. (\ref{eq:scRS}). This is not the case for GUP: under this formalism, gravity is only self-complete as long as the minimum area satisfies $\beta_0\leq1/\pi$. Such condition could be understood as a constraint on GUP. This, however, is a heuristic analysis and should be complemented by a more formal treatment (c.f. \cite{isi}).

\begin{acknowledgement}
MI and JM would like to thank the generous hospitality of the Frankfurt Institute for Advanced Studies, at which this work was initiated. MI would like to thank Loyola Marymount University Honors Program for continued support. This work has been supported by the project ``Evaporation of microscopic black holes'' (PN) of the German Research Foundation (DFG), by the Helmholtz International Center for FAIR within the framework of the LOEWE program (Landesoffensive zur Entwicklung Wissenschaftlich-\"{O}konomischer Exzellenz) launched by the State of Hesse (PN), partially by the European Cooperation in Science and Technology (COST) action MP0905 ``Black Holes in a Violent Universe'' (PN).
\end{acknowledgement}

%
%
%

\end{document}